\documentclass[preprint,journal]{vgtc}            

\title{Seeing Through the Forecast Clutter: Communicating Climate Forecast Distributions with Weighted Multiple Forecast Visualizations}

\author{%
  \authororcid{Ruishi Zou}{0009-0001-3798-6833},
  \authororcid{Siyi Wu}{0000-0003-2351-0049},
  \authororcid{Racquel Fygenson}{0000-0002-0705-9000},
  \authororcid{Dakuo Wang}{0000-0001-9371-9441},
  \authororcid{Michael Correll}{0000-0001-7902-3907}, and 
  \authororcid{Lace M. Padilla}{0000-0001-9251-5279}
}

\authorfooter{
  \item
  	Ruishi Zou is with University of California, San Diego. This work was done while visiting Northeastern University. 
  	E-mail: ruzou@ucsd.edu.
  \item
  	Siyi Wu is with University of Toronto. 
  	E-mail: reyna.wu@mail.utoronto.ca.

  \item Racquel Fygenson, Dakuo Wang, Michael Correll, and Lace M. Padilla are with Northeastern University. E-mails: \{fygenson.r, d.wang, m.correll, l.padilla\}@northeastern.edu.
}

\abstract{%
Forecasts often diverge because different models make varying assumptions to account for underlying uncertainty. Readers who consume forecasts may wish to survey the shape and spread of these multiple forecasts to get a full account of the different predictions. One approach to visualizing multiple forecasts is through Confidence Interval (CI) plots. However, while the summative CI plots can communicate uncertainty of an ensemble, they obscure attributes of individual forecasts that can lead to inaccurate perceptions of the distribution of these forecasts (e.g., implying a normal distribution when non-existent). To address this challenge, we investigate the use of multiple forecast visualization (MFV) in communicating nuanced forecast distributions through two preregistered experiments using climate forecast data. In Experiment 1 (480 participants), we compared how well MFV and CI plots can represent the distribution of multiple forecasts. We found that, compared to CI plots, MFV improved participants' ability to identify the underlying distribution of forecasts and reduced the likelihood of assuming normality. Building on Experiment 1, we examined in Experiment 2 (900 participants) whether a downsampled MFV showing 9 forecasts might be able to communicate additional forecast properties using linewidth and opacity without negatively impacting distribution perception. We found that visually weighting forecasts by linewidth or opacity preserves readers' perception of the underlying distribution. We discuss how these findings suggest the use of downsampled and weighted MFV to cut through forecast clutter by aligning perceived distribution with the underlying forecast distribution, while opening up design opportunities to use weighting to communicate additional forecast attributes.
}


\teaser{
  \centering
  \includegraphics[width=0.94\columnwidth,keepaspectratio]{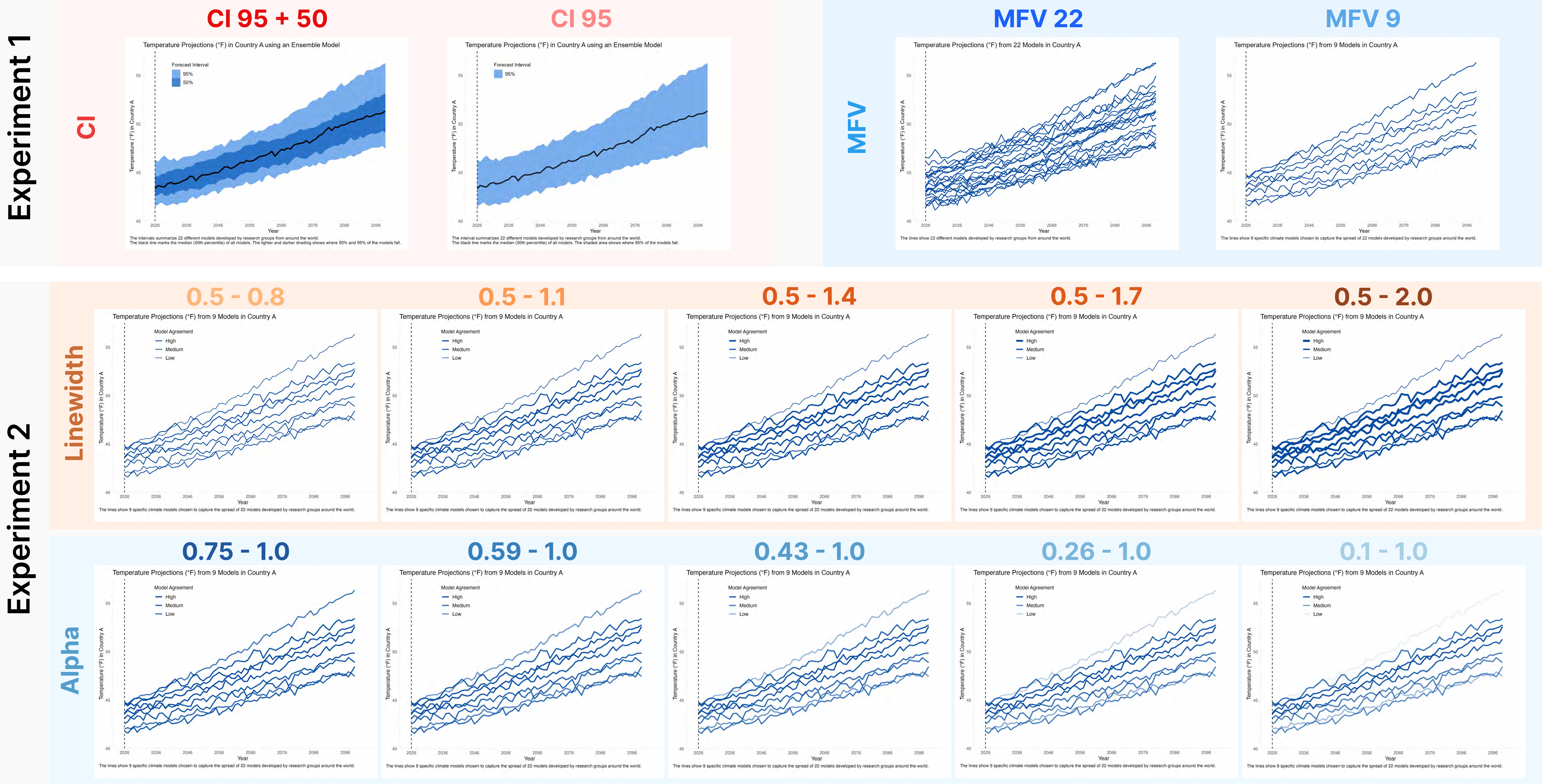}
  \caption{
        Overview of the visualization conditions in Experiments 1 and 2. Experiment 1 compares two confidence-interval visualizations (\CIfull and \CIninefive) against two multiple forecast visualizations (\MFVfull and \MFVnine). Experiment 2 evaluates weighted versions of \MFVnine using two visual encodings: \Linewidth weighting (top row) and \Alpha opacity weighting (bottom row). The examples in this figure use the Global Climate Model distribution, scenario 5-8.5. We also tested three other distributions and two other scenarios.}%
  \label{fig:teaser}
}

\graphicspath{{figs/}{figures/}{pictures/}{images/}{./}} %

\usepackage{tabu}                      %
\usepackage{booktabs}                  %
\usepackage{lipsum}                    %
\usepackage{mwe}                       %
\usepackage{ccicons}                   %
\usepackage{xcolor}                    %
\usepackage{xspace}
\usepackage{amsmath}
\usepackage{balance}

\usepackage{multirow}
\usepackage{multicol}
\usepackage{tabularx}

\definecolor{CIcolor}{HTML}{ed3a3a}
\definecolor{CI95color}{HTML}{FD8585}
\definecolor{CI9550color}{HTML}{E60D0D}
\definecolor{MFVcolor}{HTML}{299eed}
\definecolor{MFV22color}{HTML}{1C61F9}
\definecolor{MFV9color}{HTML}{5BA8E8}

\definecolor{LinewidthColor}{HTML}{CA6D37}
\definecolor{Linewidth2}{HTML}{97421D}
\definecolor{Linewidth1.7}{HTML}{BA4C1C}
\definecolor{Linewidth1.4}{HTML}{DD5A1A}
\definecolor{Linewidth1.1}{HTML}{FD9850}
\definecolor{Linewidth.8}{HTML}{FDB67A}

\definecolor{Alpha2.6}{HTML}{7AB6DA}
\definecolor{Alpha.1}{HTML}{94D0EF}
\definecolor{Alpha.75}{HTML}{21589F}
\definecolor{Alpha.43}{HTML}{559DCC}
\definecolor{Alpha}{HTML}{559DCC}
\definecolor{Alpha5.9}{HTML}{377FBC}

\newcommand{\CI}{{\textcolor{CIcolor}{CI}}\xspace}
\newcommand{\CIninefive}{{\textcolor{CI95color}{CI 95}}\xspace}
\newcommand{\CIfull}{{\textcolor{CI9550color}{CI 95+50}}\xspace}
\newcommand{\MFV}{{\textcolor{MFVcolor}{MFV}}\xspace}
\newcommand{\MFVfull}{{\textcolor{MFV22color}{MFV 22}}\xspace}
\newcommand{\MFVnine}{{\textcolor{MFV9color}{MFV 9}}\xspace}
\newcommand{\Linewidth}{{\textcolor{LinewidthColor}{Linewidth}}\xspace} \newcommand{\Linewidthtwo}{{\textcolor{Linewidth2}{Linewidth 2}}\xspace} \newcommand{\Linewidthoneseven}{{\textcolor{Linewidth1.7}{Linewidth 1.7}}\xspace} \newcommand{\Linewidthonefour}{{\textcolor{Linewidth1.4}{Linewidth 1.4}}\xspace} \newcommand{\Linewidthoneone}{{\textcolor{Linewidth1.1}{Linewidth 1.1}}\xspace} \newcommand{\Linewidtheight}{{\textcolor{Linewidth.8}{Linewidth .8}}\xspace}
\newcommand{\Alpha}{{\textcolor{Alpha}{Alpha}}\xspace}
\newcommand{\Alphatwosix}{{\textcolor{Alpha2.6}{Alpha .26}}\xspace}
\newcommand{\Alphafourthree}{{\textcolor{Alpha.43}{Alpha .43}}\xspace}
\newcommand{\Alphafivenine}{{\textcolor{Alpha5.9}{Alpha .59}}\xspace}
\newcommand{\Alphaone}{{\textcolor{Alpha.1}{Alpha .10}}\xspace}
\newcommand{\Alphasevenfive}{{\textcolor{Alpha.75}{Alpha .75}}\xspace}

\definecolor{FemaleColor}{HTML}{F1A226}
\definecolor{MaleColor}{HTML}{298C8C}

\newcommand{\colorFemale}[1]{\textcolor{FemaleColor}{#1}}
\newcommand{\colorMale}[1]{\textcolor{MaleColor}{#1}}

\usepackage{pifont}
\usepackage{graphicx}

\definecolor{greenColor}{HTML}{06746b}
\definecolor{yellowColor}{HTML}{d4a017}
\definecolor{grayColor}{HTML}{808080}

\newcommand*\inline[1]{\protect\includegraphics[height=.9em]{#1}}
\newcommand{\inlinefig}[1]{\protect\raisebox{-.2em}{\inline{wsg/#1.pdf}}}

\usepackage{mathptmx}                  %

\begin{document}

\firstsection{Introduction}
\maketitle

Readers use forecasts to form judgments and make decisions about future events. With the increase of forecast methods and combination techniques~\cite{wang2023forecast}, communicators are also increasingly showing multiple forecasts to readers in various application domains, including climate~\cite{IPCC2023SYR}, weather~\cite{mfv-windy, 2025breezyweather, noaa-mfv}, health~\cite{CDC}, and economics~\cite{croushore1993introducing, mfv-economics-example}.

To visualize multiple forecasts, designers use various encodings to communicate different forecast attributes. One approach is \textit{ensemble} visualizations, which present collections of spatiotemporal results generated from combining multiple runs or samples within a shared modeling framework~\cite{wang2019visualization}. Communicators use ensemble visualizations to express variability, sensitivity, or uncertainty from the forecast generation process, and prior work shows that ensembles can help readers reason about uncertainty~\cite{wang2019visualization, gneiting2005weather}. Meanwhile, in other scenarios, communicators may wish readers to not only understand a computed ensemble but also \textit{compare and contrast} a representative set of plausible forecast outcomes. %
Such outcomes may be created from several ensemble forecasts with meaningfully different properties, or individual forecasts with different assumptions, data inputs, or statistical methods. %
We refer to visualizations of such forecast sets that explicitly support the communication of the full forecast distribution and comparisons among forecasts as~\textit{multiple forecast visualizations} (MFV)~\cite{padilla2023multiple}.

Despite their nuanced differences, communicators often visualize ensemble and multiple forecasts through similar designs. For instance, communicators may aggregate constituents of forecast sets into a summary statistic (i.e., spread, mean), then visualize it with confidence interval plots (CIs). While CIs are compact and familiar to readers, they can be misinterpreted~\cite{correll2014error, helske2021can}. Furthermore, scholars have suggested CIs may imply normality in the underlying data distribution and obscure important features of the forecast set (e.g., clusters, multimodality)~\cite{correll2023teru,newburger2023fitting}. While CIs might be suitable in communicating attributes of a forecast ensemble, they might not afford the goal of comparing forecasts with substantively different bases for prediction (i.e., the multiple-forecast communication setting).

Additionally, individual forecasts in a set of multiple forecasts are not always epistemically equivalent. Differences in priors, assumptions, and data sources can lead to meaningful variation in forecast quality and representativeness. This information could be important to readers, yet it is rarely communicated. In climate science, for example, models differ in how well they reproduce observed patterns, and prior work has explored weighting methods based on criteria such as historical performance or independence~\cite{knutti2017climate}. Therefore, visualization designers may wish to visually distinguish each forecast to communicate meaningful differences among them to readers. Motivated by these communication needs, we ask: \textbf{\textit{How can we design visualizations for multiple forecasts that convey to readers perceptually accurate forecast distributions while affording individual comparisons?}}

To address this, we investigate downsampled and weighted MFV through two preregistered experiments (see designs and evaluation measures in \S\ref{sec:experiment-design}). In Experiment 1 (\S\ref{sec:exp1}), we compared how well MFV (full version with 22 models and a downsampled version with 9 models) and CIs enabled participants to perceive different forecast distributions in the context of global climate temperature forecasting (i.e., long-term climate projections, see~\autoref{fig:teaser}). We focused on climate visualizations as a grounded use case, which also allowed us to control for how participants' individual differences (i.e., prior beliefs about climate forecasts) might impact their perception of forecasts. We found that CIs are more likely to afford the perception of a normal distribution, and MFV improved participants' ability to distinguish between underlying distributions. In Experiment 2 (\S\ref{sec:exp2}), we examined whether visual weighting through linewidth and opacity could communicate additional forecast properties. We found that these weighted displays preserved participants' ability to perceive underlying distributions. These findings suggest that compared with equivalent CI plots, MFV can afford a more perceptually accurate foundation when the goal is to communicate both the distribution and the differences of multiple forecasts.

\smallskip
In summary, we contribute:
\smallskip

\begin{itemize}
[leftmargin=10pt, itemsep=0pt, parsep=2pt, partopsep=0pt, topsep=0pt]
\item Evidence that MFV (especially when showing 22 forecasts) can better inform readers of the underlying distribution of multiple forecasts than equivalent CI plots, which tends to imply to readers a normal underlying distribution.

\item Evaluations of two MFV visual weighting techniques (alpha and linewidth) for sampled MFV with 9 forecasts, showing that encoding density-inferred weights did not negatively impact perception of forecast distributions.

\item Evidence that readers tend to perceive a narrower range of plausible outcomes when viewing CIs compared to MFV, and low alpha encoding of a forecast also narrows perception of the forecast range.

\end{itemize}

\section{Related Work}

To situate our study, we first review prior research on visualizing forecast distributions (\S\ref{subsec:rw:uncertainty-distribution}). We then focus on works related to MFV and connect them with ensemble visualizations (\S\ref{subsec:rw:mfv}). Lastly, we introduce climate model visualization as the context of our experiment (\S\ref{subsec:rw:climate-vis}).

\subsection{Visualizing Forecast Distributions}
\label{subsec:rw:uncertainty-distribution}

Designers often characterize forecasts as distributions of plausible outcomes, which can be visualized using numerous approaches, as reviewed by Potter et al.~\cite{potter2009ensemblevis}. These approaches can be broadly organized according to their level of aggregation. A less aggregated approach displays each forecast member separately from the complete ensemble. Such an approach can be seen in use cases such as ensemble weather forecasts~\cite{ma2019interactive}, climate forecasts~\cite{dettinger2005climatechange}, and hurricane-track forecasts~\cite{liu2017uncertainty}. However, when many forecast members are displayed, they often become intertwined (i.e., ``spaghetti plots'')~\cite{dettinger2005climatechange}, making the chart visually cluttered. %
Other techniques prevent the issue through animation or sampling. For instance, Hypothetical Outcome Plots (HOPs)~\cite{kale2019hypothetical,hullman2015hypothetical} address this issue by animating ensemble members, while others reduce clutter by sampling individual outcomes~\cite{zou2026striking,liu2017uncertainty}.

More aggregated approaches use statistical methods to summarize ensemble members. Some preserved distributional information by visualizing the probability density function and/or the cumulative distribution function. These examples include probability density and cumulative distribution function plots~\cite{potter2012interactive}, violin plots, gradient plots~\cite{correll2014error}, and discretized variants such as quantile dotplots~\cite{kay2016when,fernandes2018uncertainty}, stripe plots~\cite{kay2016when}, and croissant plots~\cite{fygenson2026croissant}. Although effective for communicating the overall distribution, these visualizations of probability functions no longer preserve individual ensemble members. The most aggregated approaches represent forecast distributions by only visualizing statistical summaries (e.g., mean, median, quantiles) through intervals, box plots, or related summary marks~\cite{correll2014error,belia2005researchers,potter2010visualizing,sarma2025more,hofman2020how}. In addition to the examples that are clear in the spectrum, research has also explored hybrid approaches that combine visualizing summaries with individual ensemble members. Those examples include raindrop plots~\cite{barrowman2003raindrop}, bean plots~\cite{kampstra2008beanplot}, and contour box plots~\cite{mirzargar2014curve}.

The previously reviewed techniques motivate our comparison because they differ in the extent to which they aggregate the underlying forecast set. Lower-aggregation designs preserve individual forecasts and may better reveal distributional structure, whereas higher-aggregation designs reduce clutter but may obscure features such as clustering or multimodality. We selected two less-aggregated MFV designs, showing 9 or 22 individual forecasts (\MFVnine, \MFVfull), and two more-aggregated confidence-interval designs (\CIfull, \CIninefive), to test how aggregation level shapes readers' perception. In the following, we describe visualizing multiple forecasts as our communication scenario.

\subsection{Multiple Forecast Visualization}
\label{subsec:rw:mfv}

Despite the extensive methods for visualizing forecast distributions, prior work suggests that the effectiveness of communicating forecast uncertainty depends on the communication goal, task context, and user needs~\cite{kay2016when}. For instance, one potential setting is when forecasts come from fundamentally different predictive bases, and accurately conveying all potential outcomes is important. Such settings may occur when showing a range of possible climate~\cite{IPCC2023SYR} or weather~\cite{mfv-windy, 2025breezyweather, noaa-mfv} pathways, informing epidemiological forecasts~\cite{CDC}, and conveying potential future economic outlooks~\cite{croushore1993introducing, mfv-economics-example}. We describe visualizations that preserve forecast-level identity in these settings as MFV displays.

Because MFV does not refer to a specific visualization design, they can be displayed at varying levels of aggregation~\cite{zou2026striking}. Prior research on MFV investigated how MFV of single-variate time-series forecasts~\cite{padilla2022impact, padilla2023multiple, zou2026striking} and geospatial time-series forecasts~\cite{padilla2017effects, liu2019visualizing} impact outcomes associated with visualization reading, such as judgment performance or trust. 
Researchers have also investigated the strategies readers apply when facing simulated decision-making tasks using MFV~\cite{sarma2025more, padilla2026examininga}.

Despite prior efforts in evaluating MFV reading outcome and interpretation strategies, less is known about how they shape readers' \textit{perceptions} of the underlying forecast distribution. We argue that understanding readers' perceptions could allow us to better explain readers' strategies approaching an MFV, thus complementing prior work by demonstrating MFV's affordance. Additionally, because downstream judgments depend, in part, on whether readers perceive the intended uncertainty, we argue that to improve uncertainty communication, it is also necessary to understand readers' perception of multiple forecasts.

\subsection{Visualizing Climate Model Forecasts}
\label{subsec:rw:climate-vis}

Climate visualization has a substantial history within and adjacent to the visualization community, with recent efforts increasingly coordinated through venues such as the \textit{Viz4Climate} workshop at IEEE VIS~\cite{bach2024ieee}. In climate science, forecasts are sometimes generated using multiple Global Climate Models (GCMs), resulting in simulations that differ due to variations in model structure and assumptions~\cite{khan2025global, weart2010development}. Climate scientists then consider those multiple models holistically to account for the uncertainties of individual forecasts~\cite{tebaldi2007use,knutti2010challenges}.

In climate visualization practice, multiple climate models are often summarized before being visualized~\cite{reichler2008how,knutti2010challenges}. Visualizations of multiple climate forecasts may use an arithmetic mean as the central estimate, accompanied by percentile-based uncertainty bands (e.g., 5-95\% or 10-90\%) to encode the forecasts' spread~\cite{stephens2012communicating} (i.e., CI plots). Such summaries can be appropriate when the forecast set can be interpreted as a single distribution around a central tendency. However, they may be less suitable when the forecasts differ (e.g., when the forecasts are not equally credible, are not independent, or when comparison among forecasts is important). For instance, climate forecasts have increasingly used \textit{weighted forecast methods}, which assign different weights to individual models based on their historical performance or independence~\cite{knutti2017climate,abramowitz2019esd}. In such a use case, CIs might obscure differences across forecasts. These nuances motivate the use of visualization methods that can support comparisons among forecasts.%

Similarities between the challenges of visualizing multiple forecasts and interpreting multiple climate models motivated us to use climate forecast data in our experimental investigation of MFV. To begin this research, we first conducted a formative interview (\S\ref{sec:formative}) to connect recommendations from visualization literature to the practical needs of climate experts. We then used GCM forecast data to create our stimuli and run our experiments, before discussing our findings and their implications beyond visualizing global climate forecasts.

\section{Needs Gathering: Expert Interviews}
\label{sec:formative}

To gather the needs of communicating multiple forecasts, we conducted semi-structured interviews with three domain experts who regularly work with multiple models in climate forecasting or related fields. Using a seven-question protocol, we asked experts about their forecast experiences, when and why they communicate multiple models to the public, which forecast attributes are important, how they select or prioritize models, and common misunderstandings. We asked follow-up questions based on their responses. Each interview lasted about 30 minutes. Given the formative nature of the study, we conducted a content analysis to surface needs for visualizing multiple forecasts. The study was ethically reviewed and conducted under approved procedures by the last author's Institutional Review Board (IRB, IRB\#23-07-18). Participants provided consent before entering the study.

Across all interviews, experts described consulting multiple forecasts as a common part of real-world decision-making. They emphasized that different models, providers, and assumptions often coexist, requiring people to interpret and compare multiple sources rather than rely on a single ``best'' forecast. They suggested that uncertainty, intensity, trustworthiness, reputation, and spatial sensitivity are potential metrics to communicate about each forecast. They also stressed that multiple forecast communication is important for both experts and the general public. For instance, one expert noted ``\textit{many different groups, organizations, [and] individuals}'' rely on multiple-forecast information, while another pointed out how ordinary decisions such as ``\textit{should I go to the park today}'' or ``\textit{meet my mom for lunch outside}'' can be impacted by multiple forecast communication (e.g., weather forecasts). Finally, two experts suggested that misunderstandings are especially likely when forecasts are collapsed into a single summary. One expert noted that when people see ``\textit{a nominal value with some reported uncertainty,}'' they often do not know ``\textit{is this the full range of results}'' or ``\textit{some statistical summary}.'' Another expert observed that readers often focus on a single scenario because they ``\textit{only have time to look at one.}'' 

In summary, the concern around summaries and the need to convey differences among forecast sources motivated our subsequent stimulus design. Specifically, we compared designs that aggregated multiple forecasts to different extents and measured how those representations reflected forecast distributions. We also tested whether visual weighting could encode additional forecast-level information without disrupting perception of the overall forecast distribution.

\section{Method}
\label{sec:experiment-design}

\begin{figure*}
    \centering
    \includegraphics[width=0.94\linewidth]{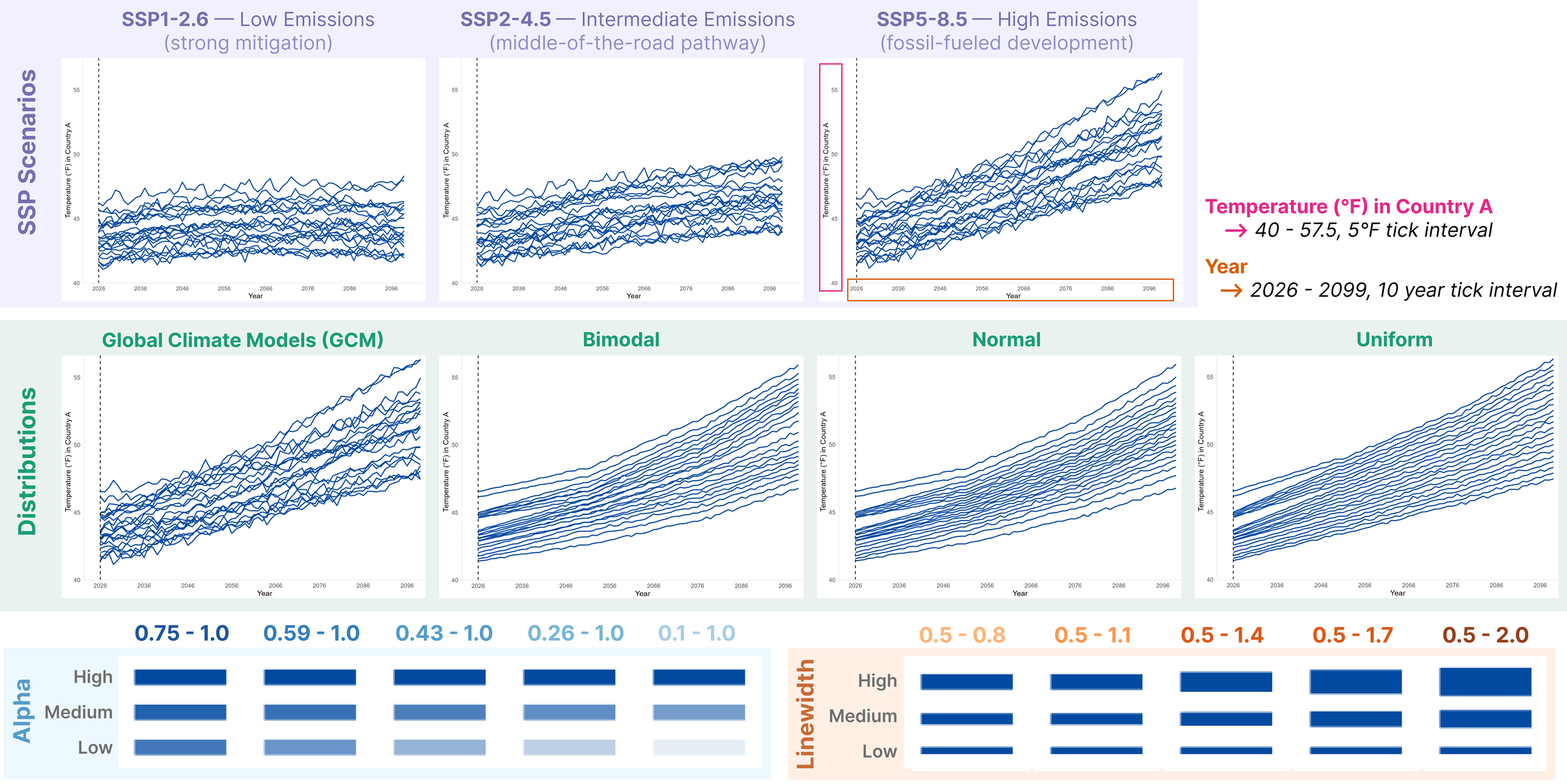}
    \caption{Illustration of the three SSP Scenarios (SSP1-2.6, SSP2-4.5, SSP5-8.5), Data Distributions (SSP5-8.5, \MFVfull example), and Visual Channels (\Alpha and \Linewidth) used in Experiments 1 and 2.}
    \vspace{-8px}
    \label{fig:scenario-distribution}
\end{figure*}

We conducted two preregistered experiments. The first aimed to compare how readers perceive \CI versus \MFV, and the second investigated how variations in visual encodings (i.e., \Linewidth and \Alpha) affect readers' visual perception of the distribution of multiple climate forecasts. In this section, we first introduce our experimental design, including the scenarios (\S\ref{subsec:scenarios}), data distributions (\S\ref{subsec:data-shapes}), stimuli (\S\ref{subsec:stimuli}), study design (\S\ref{subsec:study-design}), procedures, tasks, response options (\S\ref{subsec:procedure-tasks}), and participants (\S\ref{subsec:participants}) used in both experiments. Then, we outline our preregistered hypothesis (\S\ref{subsec:hypothesis}) and analysis method (\S\ref{subsec:analysis}).

\subsection{Climate Model Selection and Scenarios}
\label{subsec:scenarios}
We use 22 vetted GCMs from Canada's Coupled Model Intercomparison Project (CMIP6)~\cite{canada_cmip6_list_2019}. The Canadian government typically relies on 24 models for this project~\cite{ClimateDataCA}, of which we obtained 22 from the sources~\cite{GCM_data}. These models provide a realistic, externally motivated test bed for evaluating visualization techniques for multiple forecasts. We downloaded forecasts from the CMIP6 archive through the Earth System Grid Federation MetaGrid interface~\cite{ESGFMetaGridDKRZ}. We selected projections from three Shared Socioeconomic Pathway (SSP) scenarios, SSP1-2.6, SSP2-4.5, and SSP5-8.5 (\autoref{fig:scenario-distribution} SSP Scenarios). These three scenarios are a subset of the five scenarios commonly used by the IPCC~\cite{IPCC2023SYR}, selected to represent low-, intermediate-, and high-emissions futures and to span a plausible range of future climate outcomes under different levels of emissions and climate action. We treated the SSP scenarios as a \textbf{repeated measure} to test whether the effects we observe generalized across substantively different climate futures. We also selected the scenarios from the GCM data instead of using other datasets to prevent introducing confounds caused by inherent dataset biases.

\subsection{Data Distributions}
\label{subsec:data-shapes}
To evaluate whether participants could accurately perceive different forecast distributions, we generated three controlled distributions from the realistic \textbf{GCM} data, including \textbf{Uniform}, \textbf{Normal}, and \textbf{Bimodal} (\autoref{fig:scenario-distribution} Distributions). We selected these distributions to capture distinct but plausible forecast structures (evenly dispersed, symmetric, and clustered/multi-modal). Together with the original GCM distribution, they allowed us to test whether visualization effects generalized across distributional shapes rather than being specific to one forecast set.

We created the three simulated shapes by modifying the distribution of GCM data. To maintain the GCM's general trend, we selected four anchor years between the current and forecast horizons (2026, 2050, 2075, and 2099). For the first anchor year (2026), we used the original data points from the GCM. For subsequent anchor years, we generated distributions based on summary statistics from the GCM data. For the Normal and Bimodal shapes, we preserved the GCM's original median to maintain the overall forecast trajectory and adjusted the standard deviation to ensure a similar forecast spread. For the Uniform shape, we defined the distribution range using the original GCM's minimum and maximum values. Additionally, for the Bimodal shape, we applied a heuristically determined shift parameter to ensure the two modes were visually distinguishable. We used linear interpolation to connect the forecasts between the anchor years and added Gaussian noise to simulate fluctuations. We document the parameters used to generate Uniform, Normal, and Bimodal distributions in~\autoref{appendix:data-dist-parameters}.

\subsection{Stimuli Visualizations}
\label{subsec:stimuli}

We created four visualization types: two types of \CI: \CIfull, \CIninefive, and two types of \MFV: \MFVfull, \MFVnine (see \autoref{fig:teaser},~Experiment 1). The visualization type is our primary manipulation in both experiments. We based the visualization captions on educational information from \texttt{ClimateData.ca} on how to understand climate forecasts~\cite{ClimateDataCA}.

We designed the two \CI visualizations as examples of summary-based representations. Both CIs use a black line to display the median curve (connecting the median at each time point) and a blue band to display the intervals. \CIfull shows both a 95\% and a 50\% confidence interval, whereas \CIninefive shows only a 95\% confidence interval.

\MFVfull displays all 22 independent forecasts for each given data distribution, whereas \MFVnine downsamples \MFVfull to a subset of 9 representative forecasts, which is within the recommended number of forecasts for \MFV~\cite{padilla2023multiple}. Our goal in testing \MFVnine was to determine whether participants could still perceive the underlying distribution after downsampling. We therefore derived \MFVnine from \MFVfull to directly assess whether the loss of forecast trajectories affected distribution perception. To ensure \MFVnine captures the full spread of \MFVfull, we applied a percentile-based sampling method: 1) we defined 9 equally-spaced percentiles (from 0\% to 100\% in 12.5\% increments) across the full time series, then 2) we calculated the normalized distance (i.e., the sum of absolute differences after min-max normalization at all time points) of each forecast to each percentile, and 3) we selected the model closest to each percentile as its ``representative.'' If the same forecast was selected for multiple percentile levels, the procedure terminates and flags the case for manual resolution by the authors. We detected no duplicate assignments while creating our stimulus. 

In addition to the four visualization types, we experimented with variants of \MFVnine in Experiment 2 (\autoref{fig:teaser},~Experiment 2). Our goal was to test whether MFV can encode model-level differences while preserving perception of the overall forecast distribution. While expert interviews suggested many potentially useful attributes (e.g., trustworthiness, reputation), estimating these quantities requires domain-specific assumptions and introduces additional confounds (e.g., individual differences in the belief in these metrics) that are beyond the scope of the study. Therefore, we chose a data-driven attribute, ``consensus'', as a proxy for the weightings. Consensus captures how well each displayed forecast represents the full forecast set of forecasts. For example, in our weighting process, we interpret forecasts with more adjacent forecasts as more representative of the collective forecast set and should therefore be more visually emphasized, whereas isolated forecasts should receive less emphasis. Because these consensus weights are derived from the same forecast distribution being visualized, they are not independent measures of model quality. We used them only to provide a controlled test case to assess whether visual weighting can convey forecast-level information without disrupting readers' perception of the overall forecast distribution.

To compute consensus weights and map them to \MFVnine, we computed a kernel density estimate (using R's \texttt{density()} function with bandwidth \texttt{nrd0} and 512 grid points) of all forecasts at the forecast horizon (i.e., year 2099), applied a power transformation with exponent 1.6 to accentuate differences, then linearly rescaled the transformed values to a range of 0.5 to 1.4. These weights were then used directly for the \Linewidth condition and linearly mapped to an opacity range of 0.1 to 1 for the \Alpha condition (\autoref{fig:scenario-distribution} Alpha, Linewidth). 

Specifically, we considered \textbf{alpha} (opacity) and \textbf{linewidth} because they are ``magnitude channels'' and thus appropriate for displaying weight values~\cite{munzner2014visualization, padilla2021uncertainty}. While other channels, such as color or dashing may also convey weighting, alpha and linewidth are parsimonious additions to MFV~\cite{sterzik2024perception}. To determine linewidth levels, we tested a range of widths between the lower bound of 0.5 and one of five higher bounds (0.8, 1.1, 1.4, 1.7, and 2.0). We tested alpha levels ranging from five lower bounds (0.10, 0.26, 0.43, 0.59, and 0.75) to an upper bound of 1.0 (i.e., full alpha), as shown in~\autoref{fig:teaser}. We selected those levels based on previous psychophysical, perceptual modeling, and visualization research~\cite{stone2014engineering,fechner1948elements,sterzik2024perception} (see rationale in \autoref{appendix:alpha-linewidth-levels}).

While the sampling and weighting methods we used are simplifications of those used in real-world practice, they provide a deterministic and interpretable approach to modify forecast attributes, allowing us to focus on investigating the perceptual effects of sampling and weighting.

\subsection{Study Design}
\label{subsec:study-design}
\begin{figure*}[t]
\centering
\includegraphics[width=1\textwidth,keepaspectratio]{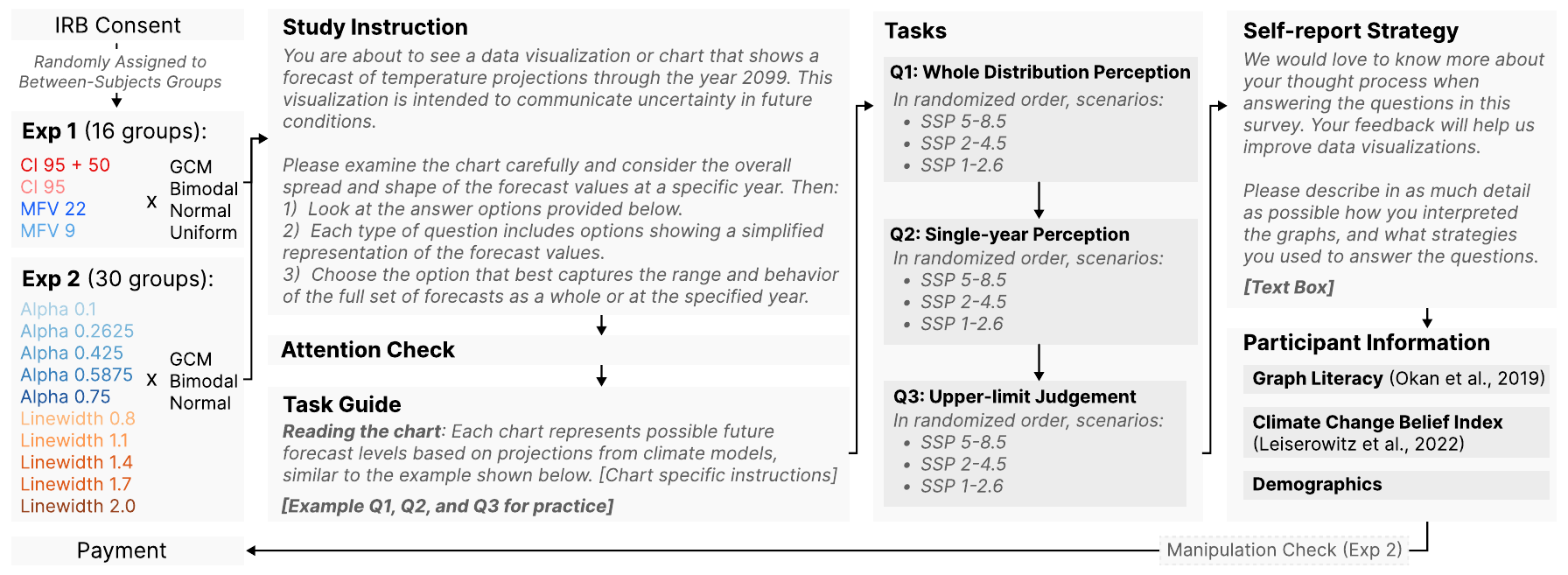}
\caption{
Overview of the experimental procedure for Experiments 1 and 2. }
\label{fig:exp-flowchart}
\vspace{-8px}
\end{figure*}

We employed a mixed between- and within-subject design for both Experiments (\autoref{fig:exp-flowchart}). Experiment 1 employed a 4 $\times$ 4 $\times$ 3 mixed factorial design, with Visualization Type (\MFVfull, \MFVnine, \CIfull, \CIninefive) and Data Distribution (Uniform, Normal, Bimodal, GCM) as between-subject factors. The within-subject factor was SSP scenario (SSP1-2.6, SSP2-4.5, SSP5-8.5). Each participant evaluated forecasts under all three SSP scenarios in randomized order within their assigned visualization and distribution conditions.

Experiment 2 adopted the same mixed factorial structure as Experiment~1, but with a modified between-subject factor. Experiment 2 employed an 11 $\times$ 3 $\times$ 3 mixed factorial design, with the between-subject factors Weighting Condition (11 levels: five linewidth-weighting levels, five alpha-weighting levels, and a no-weighting baseline) and Data Distribution (Bimodal, Normal, GCM). These factors were fully crossed, yielding 33 between-subject conditions. Similar to Experiment 1, SSP scenarios (SSP1-2.6, SSP2-4.5, SSP5-8.5) were manipulated within-subjects and presented in randomized order.

\subsection{Procedure, Tasks, and Response Options}
\label{subsec:procedure-tasks}

\begin{figure}[t]
\centering
\includegraphics[width=0.86\columnwidth,keepaspectratio]{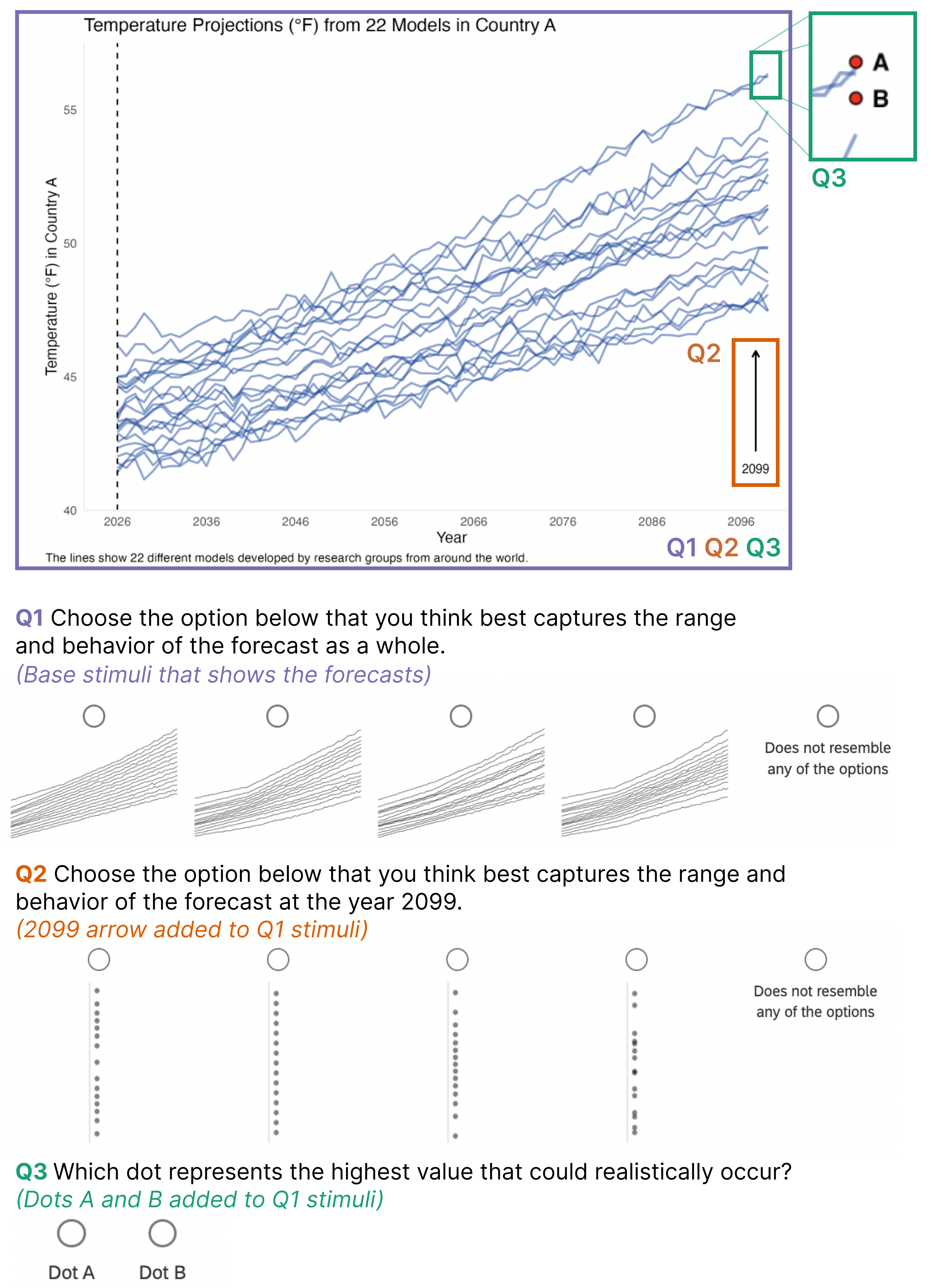}
\caption{
Illustration of the three tasks for Experiments 1 and 2 (SSP-8.5 scenario). The three questions used the same base stimuli as Q1 (purple), with additional visual elements for Q2 and Q3 (orange, green).
}
\vspace{-8px}
\label{fig:question-examples}
\end{figure}

Prior to completing the studies, participants provided informed consent in accordance with the approved procedures from the last author's institute's IRB (IRB\#23-07-18). Participants were randomly assigned to the between-subject conditions. Within each task block, participants answered the same question for all three SSP scenarios, with the scenario order fully randomized. Within each scenario, participants completed three task blocks containing three questions (\autoref{fig:question-examples}):

\smallskip
\noindent\textbf{Q1: Whole-distribution perception.} 
Participants selected one of five (Uniform, Normal, Bimodal, GCM, and non-above) candidate options that best matched the overall forecast stimulus. To generate the response options for this question, we used a similar approach to creating \MFVnine, but 1) preserved 15 forecasts instead of 9, and 2) reduced noise to reduce visual clutter (\autoref{fig:question-examples}, Q1). We also cropped the chart axes to focus readers' attention on the shapes of the candidate distributions. We selected 15 forecasts because this number falls between 9 and 22, thereby avoiding bias toward either the smaller or larger MFV sets in the response options. We also included an option to select ``Does not resemble any of the options.''

\smallskip
\noindent\textbf{Q2: Single-year perception (2099).}
Participants selected the distribution that best matched the stimulus's distribution at year 2099. The second response option was designed to assess whether readers can perceive forecast distribution at a single time point (\autoref{fig:question-examples}, Q2). To generate response options, we used the 15 forecasts from Q1's response options, only showing the responses' distribution at the forecast horizon via a strip plot. Similar to Q1, we added a ``none-above'' option.

\smallskip
\noindent\textbf{Q3: Upper-limit judgment.}
Participants chose between two high-value points (A vs. B, \autoref{fig:question-examples}, Q3) representing plausible extreme outcomes and indicated which represented a realistic upper bound. We positioned point A on or above the highest value in the dataset, while point B was slightly lower than the largest value.

\smallskip
The three questions were designed to collectively capture readers' perception of multiple forecast distributions from complementary perspectives. Specifically, Q1 measured readers' perception of the overall shape of the forecasts. Q2 investigated whether readers could perceive the distribution at a single specified time point. We included Q2 as a complementary measure to Q1, providing an alternative way to assess perceived forecast distribution while reducing the likelihood that our findings were driven by a specific visual design of the response options.

Q3 investigated participants' perceived extent of the displayed forecast set. Thus, Q3 complemented Q1 and Q2 by measuring whether different visualization designs led readers to place the upper edge of the forecast range closer to the most extreme displayed forecast or to the interior of the distribution. This allowed us to analyze how visualization design shaped readers' perception of the displayed forecast range.

\smallskip

After completing these tasks, participants completed measures of visualization literacy~\cite{okan2019using}, the Climate Change Belief Index~\cite{ClimateSurvey, ballew2019climate}, and demographic questions. 
We chose to use the short-form test (4 items) over more comprehensive tests such as VLAT (53 items)~\cite{lee2017vlat} or mini-VLAT (12 items)~\cite{pandey2023minivlat} to prioritize evaluation brevity over comprehensiveness. We also included three relevant items from the Climate Change Belief-Attribution-Worry (BAW) Index based on Leiserowitz et al.~\cite{ClimateSurvey}, as participants' prior beliefs about climate change may shape how they interpret forecast information. We treated graph literacy and climate change belief as \textbf{covariates} in our analysis.

Experiment 2's procedure and tasks were identical to Experiment~1's. One addition {was} a manipulation check at the end to assess participants' ability to distinguish linewidth and alpha differences.

\subsection{Participants}
\label{subsec:participants}

We recruited participants via Prolific~\cite{Prolific2026}. Participants were at least 18 years old, resided in the United States, fluent in English, and completed the study on a computer. Given that climate-related beliefs may vary across demographics, we used Prolific's U.S. representative sampling option to recruit a demographically balanced sample matched to the U.S. population by age, sex, and ethnicity based on the 2021 census~\cite{ProlificRepresentativeSamples}. We conducted the study with a sample that approximated the U.S. population because our expert interviews suggested that understanding climate forecasts is important for both experts and the public. 

We recruited 480 participants for Experiment~1 and 900 participants for Experiment~2 (30 per between-subjects group for both experiments). 
We excluded participants who failed the attention check or the manipulation test, resulting in final samples of 440 participants in Experiment~1 and 796 participants in Experiment~2. The baseline group of Experiment 2 (no weighting) was taken from Experiment 1.

In Experiment~1, participants were 52.5\% \colorFemale{women} (\(n = 231\)), 45.0\% \colorMale{men} (\(n = 198\)), 2.0\% non-binary (\(n = 9\)), and 0.5\% preferred not to say (\(n = 2\)) \inlinefig{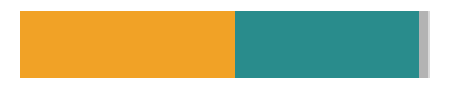}. Ages ranged from 18 to 87 years, with mean ages of 44.1 (\(SD = 16.0\)) for \colorFemale{women} \inlinefig{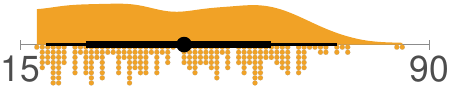} and 42.6 (\(SD = 15.2\)) for \colorMale{men} \inlinefig{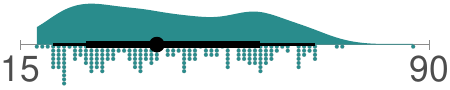}. In Experiment~2, participants were 51.6\% \colorFemale{women} (\(n = 411\)), 45.5\% \colorMale{men} (\(n = 362\)), 2.6\% non-binary (\(n = 21\)), and 0.3\% preferred not to say (\(n = 2\)) \inlinefig{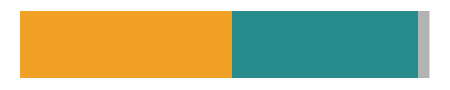}. Ages ranged from 18 to 82 years, with mean ages of 41.9 (\(SD = 15.2\)) for \colorFemale{women} \inlinefig{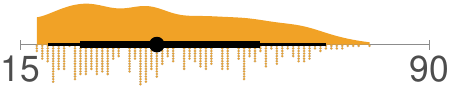} and 39.9 (\(SD = 15.0\)) for \colorMale{men} \inlinefig{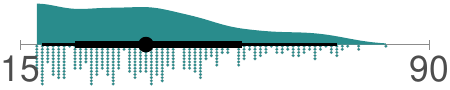}. Across both experiments, participants demonstrated moderate graph literacy~\cite{okan2019using} and high levels of belief in climate change and concern~\cite{ClimateSurvey}. The median graph literacy score was 2 of 4 \inlinefig{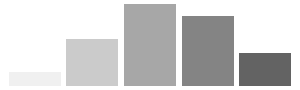}, and the median climate change worry score was 0.89 on a 0-1 scale \inlinefig{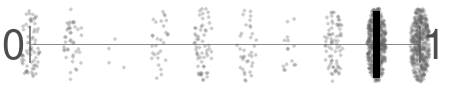}.

\subsection{Hypotheses}
\label{subsec:hypothesis}

\textbf{Exp 1: H1A-C (\MFV advantage for complex distributions).}
For the Bimodal, Uniform, and GCM distributions, we predicted that participants viewing \MFVfull visualizations would be more likely to correctly identify the underlying data distribution than participants viewing \CIninefive and \CIfull visualizations (\textbf{H1A}), and more likely to do so than participants viewing \MFVnine visualizations (\textbf{H1B}). We also predicted that participants viewing \MFVnine visualizations would outperform those viewing \CI visualizations (\textbf{H1C}). 

These hypotheses were motivated by prior work suggesting that displays of individual forecasts preserve more information about the underlying distribution than summary-based uncertainty visualizations~\cite{potter2009ensemblevis}. Additionally, our prediction that \MFVnine would outperform \CI visualizations was motivated by prior work suggesting that showing approximately nine forecasts may balance interpretability and trust~\cite{padilla2023multiple}. However, prior work has not established whether nine forecasts are sufficient to preserve perception of full distributional shape.%

\smallskip

\noindent \textbf{Exp 1: H2A-D (\CI biased towards normal assumption).}
We restricted the prior hypotheses to the Bimodal, Uniform, and GCM conditions because we expected confidence interval visualizations to imply a roughly normal distribution, as previously suggested~\cite{newburger2023fitting}. Prior work on hurricane forecast visualizations found that viewers are strongly influenced by interval boundaries while still inferring a graded likelihood distribution within and beyond the interval~\cite{ruginski2016nonexpert}. Although we are not aware of prior work directly showing that confidence intervals imply normality, we hypothesized that this is true given the shape of \CI plots. %

Thus, when the true generating distribution was not Normal, we predicted that participants viewing \CIfull (\textbf{H2A}) and \CIninefive (\textbf{H2B}) visualizations would be more likely than those viewing \MFV to select Normal response options. When the true distribution was Normal, however, we expected \CIfull (\textbf{H2C}) and \CIninefive (\textbf{H2D}) visualizations to produce rates of Normal responses similar to those of \MFV. In other words, we predicted \textit{an interaction} between visualization type and data distribution, which we describe in our interaction analysis (\S\ref{subsec:analysis}). 

\smallskip

\noindent \textbf{Exp 1: H3 (Extent heuristic).}
Across visualization types, we expected viewers to interpret the furthest visible extent of a mark as the realistic upper bound of possible outcomes. This prediction is consistent with recent work on \MFV, which shows that participants tend to place upper and lower boundaries at the visible extent of each mark~\cite{padilla2026examininga}. Because confidence interval visualizations do not always reflect the full range of the underlying model values, we predicted that participants viewing \CI would be more likely than those viewing \MFV to select lower-valued upper limit responses. H3 is relevant to Experiment 2, in which we manipulate \Alpha and \Linewidth levels and investigate how they impact readers' perception of the forecast extent. We include H3 in Experiment 1 as a baseline for Experiment 2.%

\smallskip

\noindent \textbf{Exp 2: H4 (Weighted Forecasts vs.\ No Weighting).}
We predicted that at least one weighting condition (either \Linewidth or \Alpha weighting) would improve accuracy in identifying the underlying data distribution relative to the no-weighting baseline. This prediction was motivated by the idea that visually emphasizing more highly weighted forecasts could increase the salience of the distributional structure.

\smallskip

\noindent \textbf{Exp 2: H5 (Moderate Weighting Advantage).}
We predicted moderate weighting would yield the highest accuracy in informing accurate data distribution, relative to both weak and extreme weighting levels. We reasoned that weak weighting may be insufficient to influence perception, whereas overly strong weighting may negatively affect accuracy by drawing excessive attention to only strongly weighted forecasts.

\smallskip
\noindent \textbf{Exp 2: H6 (Extent Effect).} As described in~\S\ref{subsec:stimuli}, visual weight and forecast position dual-encoded the same information, and we hypothesize that participants would be more likely to select lower-valued upper-limit responses when extreme forecasts were visually de-emphasized with lower opacity or thinner lines (both \Alpha and \Linewidth manipulations). This prediction follows from the assumption that sufficiently faint or thin extreme forecasts may contribute less to participants' perceived upper extent of plausible outcomes.

\subsection{Analysis Method}
\label{subsec:analysis}

We used multinomial \textbf{Bayesian multilevel modeling}, implemented with the R package \texttt{brms}, because our outcome variable was the response category selected in a multiple-choice question. Given this multinomial outcome structure, Bayesian categorical regression was an appropriate modeling approach.
We chose Bayesian modeling over Frequentist approaches because it estimates full posterior distributions over response probabilities, allowing us to analyze contrasts between visualization conditions in terms of both effect magnitude and uncertainty. This avoids reducing results to binary significance decisions and supports the paper's broader goal of reasoning about nuanced distributional information rather than only summary statistics.

When reporting contrasts between visualization conditions, we report $\Delta$ and Credible Intervals (CrI). Specifically, $\Delta$ is the posterior mean difference in the predicted probability between conditions (e.g., MFV-22 minus CI-95), with positive values indicating higher predicted accuracy. CrI that excludes zero indicates that the posterior distribution consistently favors one condition over the other, which roughly corresponds to the Frequentist concept of ``significance.'' 
 
\begin{figure*}[t]
\centering
\includegraphics[width=0.95\textwidth]{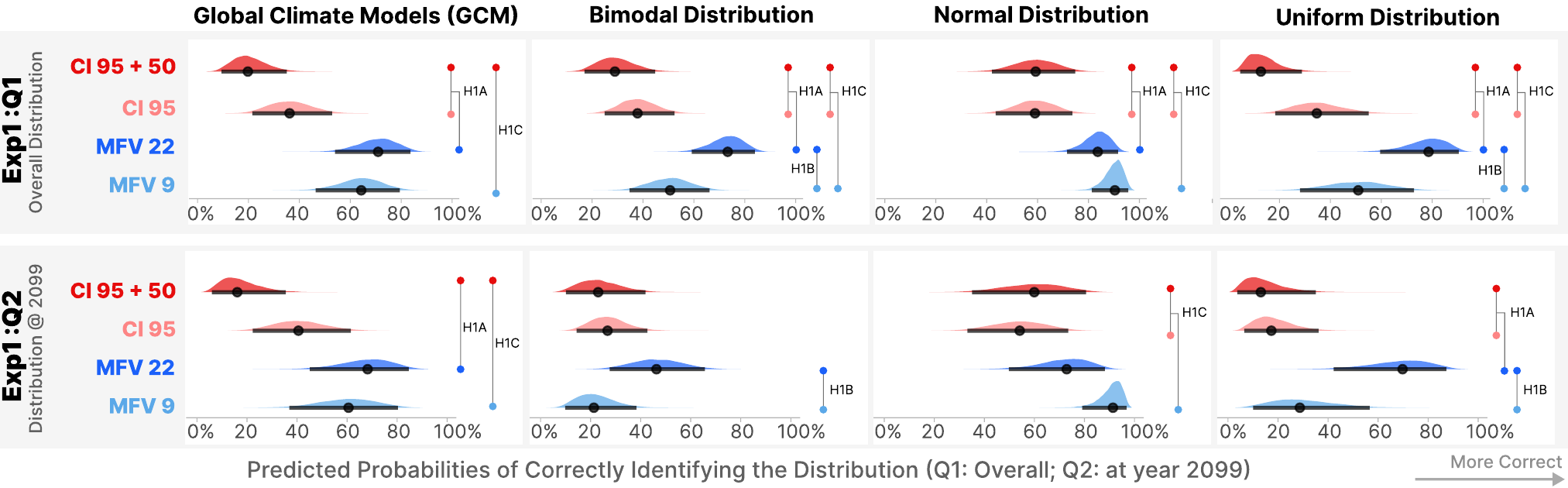}
\caption{Posterior predicted probability densities for Experiment 1 Q1 and Q2. Each row of each figure corresponds to the visualization conditions, whereas each column corresponds to the underlying data distribution. The black dot-interval represents the posterior means and the 95\% credible intervals (CrI). Hypotheses annotated on the figure represent meaningfully different contrasts (i.e., the posterior contrast 95\% CrI does not include 0).
}
\vspace{-8px}
\label{fig:exp1-results}
\end{figure*}

\paragraph{Q1 and Q2}
We model participants' responses for Q1 and Q2 using \textbf{Bayesian
multilevel categorical logistic regression}. Specifically, for each observation 
$i$, we denote the response as $Y_i \in \{\textit{None},\textit{Norm},\textit{Bi},\textit{Uni},\textit{GCM}\}$, treating \textit{None} as the baseline category. For experiment 1, the response probabilities were modeled as:
\[
Y_i \sim \text{Categorical}(\boldsymbol{\pi}_i)
\]
where $\boldsymbol{\pi}_i = (\pi_{i,\text{None}}, \pi_{i,\text{Norm}}, \pi_{i,\text{Bi}}, \pi_{i,\text{Uni}}, \pi_{i,\text{GCM}})$ is the vector of category probabilities for observation $i$, which sum to 1. For each non-baseline category $k$, we modeled the log-odds as:
\[
\begin{aligned}
\log\!\left(\frac{\pi_{ik}}{\pi_{i,\text{None}}}\right)
&=
\alpha_k
+ \beta_{1k}\,\text{cond}_i
+ \beta_{2k}\,\text{dataDist}_i\\
&\quad
+ \beta_{3k}\,(\text{cond}_i \times \text{dataDist}_i)
+ \beta_{4k}\,\text{scenario}_i \\
&\quad
+ \beta_{5k}\,\text{C\_graphLit}_i
+ \beta_{6k}\,\text{C\_BAW}_i
+ u_{j[i],k}
\end{aligned}
\]
Specifically, predictor variables include the 
visualization condition (\texttt{cond}, two \CI and two \MFV for Experiment 1), 
the underlying data distribution (\texttt{dataDist}), their interaction, 
the SSP scenario (\texttt{scenario}), 
centered visualization literacy (\texttt{C\_graphLit}), and 
centered climate change Belief-Attribution-Worry Index (\texttt{C\_BAW}). We include an interaction between \texttt{cond} and \texttt{dataDist} because our hypotheses predicted that some effects of the visualization condition would differ for Normal distributions relative to the other distributions (H1s-H2s). 

Participant-level variability was modeled with random intercepts $u_{j[i],k} \sim \text{Normal}(0, \sigma_k^2)$, where $j[i]$ indexes the participant for observation $i$, allowing each participant to have a unique baseline tendency for each response category.
We modeled participants as a random intercept because each participant completed repeated trials across SSP scenarios. Thus, observations from the same participant were not independent. We accounted for this repeated-measures structure by including the SSP scenario as a fixed effect and participant-level random intercepts, which allow each participant to have their own baseline response tendency.

Additionally, we used weakly informative priors to regularize parameter estimates and reduce the risk of overfitting, where:
\[
\begin{aligned}
\alpha_k,\beta_{pk}\sim\text{Normal}(0,1) \quad \sigma_k\sim\text{Exponential}(1)
\end{aligned}
\]

For Experiment 2, we adopted a similar modeling approach 
to Experiment 1, but simplified the fixed-effects structure. The visualization condition (\texttt{cond}: a no-weighting baseline, five levels each of \Linewidth and \Alpha) and the data distribution (\texttt{dataDist}, excluding Uniform because uniform distribution have no weighting) were included as main effects without an interaction, because we did not anticipate that the effects of \Linewidth or \Alpha would vary across data distributions. %

\paragraph{Q3}
Q3 followed a similar analysis process to Q1 and Q2, but adjusted the outcome variable and link function to fit the question type. Specifically, we used \textbf{Bayesian multilevel binomial logistic regression} to model participants' binary choice between the two extreme points (DotA: at the most extreme forecast vs. DotB: just within the most extreme forecast). Let $Y_i \in \{\text{DotA}, \text{DotB}\}$ denote the response for observation $i$, with $\text{DotB}$ treated as the baseline category. The response probability was modeled as:
$
Y_i \sim \text{Bernoulli}(\pi_i)
$
where $\pi_i$ is the probability of selecting DotA. For each observation, we define the logit as
$
\log\!\left(\frac{\pi_i}{1 - \pi_i}\right)
$
which is a function of visualization condition, data distribution, their interaction (only for Experiment 1), scenario, centered visualization literacy, centered climate change beliefs, and participant-level random intercepts, similar to Q1 and Q2. Q3 also used the same weakly informative priors as Q1 and Q2.

In the following sections, we report our analysis results. We document a summary of all results in \autoref{appendix:hypo-outcome}.

\section{Results: Experiment 1}
\label{sec:exp1}

\subsection{\textbf{H1A-C}: MFV advantage for complex distributions}

\textbf{H1A: \MFVfull visualizations produce higher accuracy in identifying the true generating distribution than \CI-based visualizations.} 
As shown in the top panel (Q1) and bottom panel (Q2) of \autoref{fig:exp1-results} (see H1A annotations), \MFVfull generally produced higher predicted probabilities of correct responses than the \CI-based visualizations. In Q1, these advantages were consistent across all four generating distributions with the largest effect observed for the Uniform distribution when comparing \MFVfull to \CIfull ($\Delta = 0.64$, 95\% CrI [0.43, 0.81]). In Q2, the more accurate perception of \MFVfull remained strongest for the Uniform distribution relative to \CIninefive; and for the GCM distribution when comparing \MFVfull to \CIfull (\autoref{fig:exp1-results}, Experiment 2, H1A annotation). Effects for the Bimodal, GCM relative to \CIninefive, and Normal distributions in Q2 were in the same positive direction but less certain, with credible intervals overlapping zero. 

These results \textbf{support H1A}, with most contrasts favoring \MFVfull over \CI-based visualizations and no contrasts contradicting the hypothesis. This indicates that \MFVfull visualizations may improve participants' ability to identify the underlying distribution relative to \CI-based visualizations, although the strength varied across comparisons.

\smallskip
\noindent
\textbf{H1B: \MFVfull visualizations will produce higher accuracy than \MFVnine visualizations.}
Comparisons between \MFVfull and \MFVnine showed a similar pattern across Q1 (top panel) and Q2 (bottom panel) in \autoref{fig:exp1-results}. In both questions, \MFVfull produced higher predicted probabilities of correct responses for the Uniform and Bimodal distributions, whereas differences for the GCM distribution were smaller and less certain. For the Normal distribution, differences were uncertain in Q1 and trended in the opposite direction in Q2, with \MFVnine outperforming \MFVfull, although the credible interval still included zero ($\Delta = -0.19$, 95\% CrI [-0.41, 0.00]). Despite the inverted direction, both \MFVnine and \MFVfull achieved good task accuracy for the Normal distribution. These results provide \textbf{limited support for H1B}.

\smallskip
\noindent
\textbf{H1C: \MFVnine visualizations will produce higher accuracy than \CI-based visualizations.}
\MFVnine generally produced higher predicted probabilities of correct responses than the \CI-based visualizations (annotated with H1C in~\autoref{fig:exp1-results}). We observed the clearest advantages for the Normal distribution across both questions. We also observed credible positive differences comparing \MFVnine with \CI-based visualizations for the GCM distribution in Q1, as well as for the Uniform distribution relative to \CIfull in Q1. The rest of the comparison followed the same direction but was less certain, with the only exceptions being the Bimodal distribution in Q2 (\MFVnine vs. \CIfull: $\Delta = -0.02$, 95\% CrI [-0.22, 0.17]; \MFVnine vs. \CIninefive: $\Delta = -0.05$, 95\% CrI [-0.22, 0.11]). Overall, these results provide \textbf{moderate support for H1C}.

\begin{figure*}[t]
\centering
\includegraphics[width=.75\textwidth,keepaspectratio]{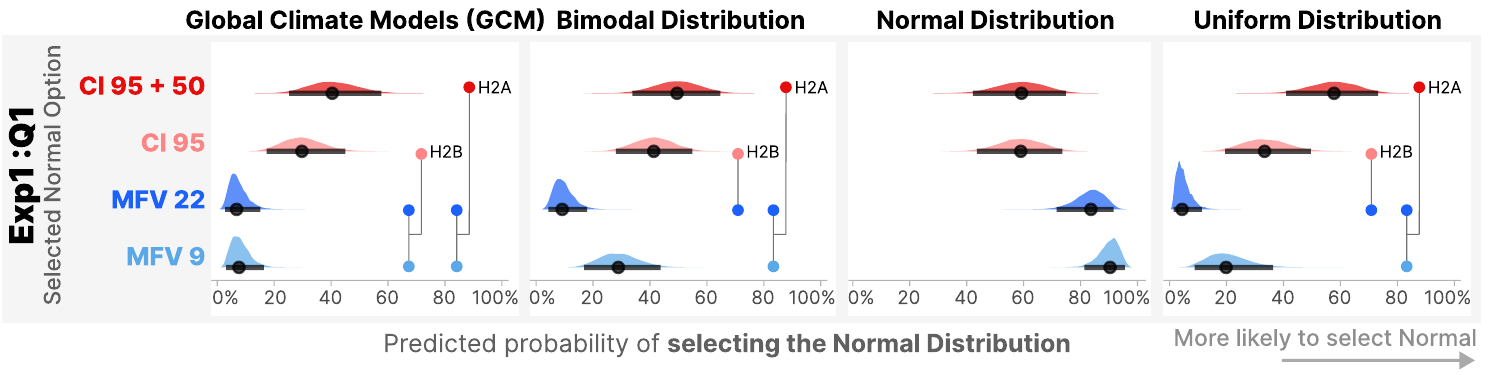}
\caption{
Posterior predicted probability densities of selecting the Normal distribution across data distributions and conditions in Exp~1 Question 1.}
\vspace{-8px}
\label{fig:exp1-normal-response}
\end{figure*}

\subsection{\textbf{H2A-D}: CI biased towards normal assumption} 

\textbf{H2A-B: When the underlying distributions are \textit{not} Normal, \CIfull (H2A) and \CIninefive (H2B) visualizations will elicit more Normal responses compared to \MFV visualizations.} 
As illustrated in \autoref{fig:exp1-normal-response}, participants are more likely to select the Normal response viewing \CI than \MFV. In Q1, \CIfull produced higher rates of Normal responses than both \MFV conditions for the Bimodal, GCM, and Uniform distributions (\autoref{fig:exp1-normal-response} H2A). \CIninefive showed a similar pattern, though differences versus \MFVnine were smaller and less consistent (\autoref{fig:exp1-normal-response} H2B). In Q2, the same overall pattern held but was weaker and less consistent. \CIfull exceeded \MFVfull for Bimodal, GCM, and Uniform distributions, while \CIninefive exceeded \MFVfull for Bimodal and Uniform but not clearly for GCM ($\Delta = 0.05$, 95\% CrI [-0.11, 0.20]). 
Meanwhile, differences between \CIninefive and \MFVnine remained positive but uncertain for the Uniform and GCM distributions, whereas the Bimodal contrast trended in the opposite direction ($\Delta = -0.07$, 95\% CrI [-0.26, 0.10]). Overall, these results \textbf{support H2A} (\CIfull) but provide only \textbf{moderate support for H2B} (\CIninefive).

\smallskip
\noindent
\textbf{H2C-D: When the underlying distributions are Normal, \CIfull (H2C) and \CIninefive (H2D) will elicit similar rates of selecting Normal responses as \MFV.} 
When the distributions were Normal, \MFV conditions elicited more correct Normal responses than \CI conditions (\autoref{fig:exp1-normal-response}, Normal Distribution). 
Specifically, in Q1, both \CI types produced fewer Normal responses than both \MFV conditions. In Q2, the same direction held but was weaker. \CI conditions still produced fewer Normal responses than \MFVnine, while differences versus \MFVfull were smaller. While these results \textbf{do not support H2C-D}, they show \MFV's potential to convey the correct perception of normal distributions when the data distribution is normal.

\begin{figure}[h]
\centering
\includegraphics[width=0.75\columnwidth]{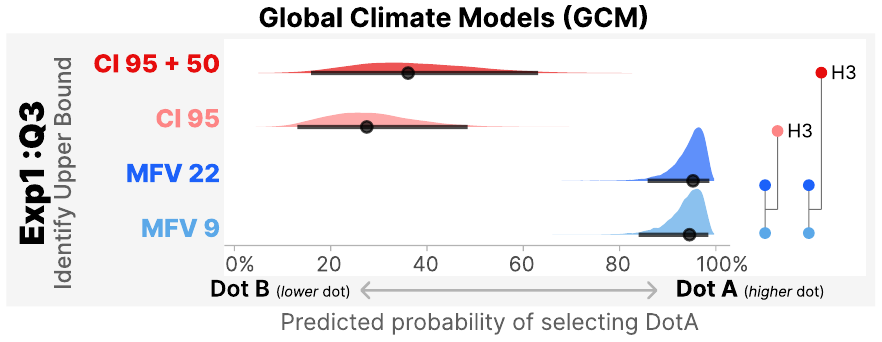}
\caption{
Posterior predicted probability densities of selecting the higher extreme value (Dot A) in Experiment~1 (GCM distribution). The same effect applies to all other data distributions.}%
\vspace{-8px}
\label{fig:exp1-q3-results}
\end{figure}

\subsection{\textbf{H3: Extent Heuristics}}
\textbf{H3: \CI-based visualizations will be more likely than \MFV to elicit lower-valued upper-limit responses.} 
Across all distributions, \MFV conditions produced much higher probabilities of selecting Dot A than \CI conditions  (\autoref{fig:exp1-q3-results}). These results \textbf{support H3}, indicating that viewers interpret the furthest visible extent of a visual mark as the plausible upper bound of outcomes. Consequently, participants viewing \CI visualizations were more likely to select lower-valued upper-limit responses than participants viewing \MFV visualizations.

\section{Results: Experiment 2}
\label{sec:exp2}

\begin{figure*}[t]
\centering
\includegraphics[width=1\textwidth,keepaspectratio]{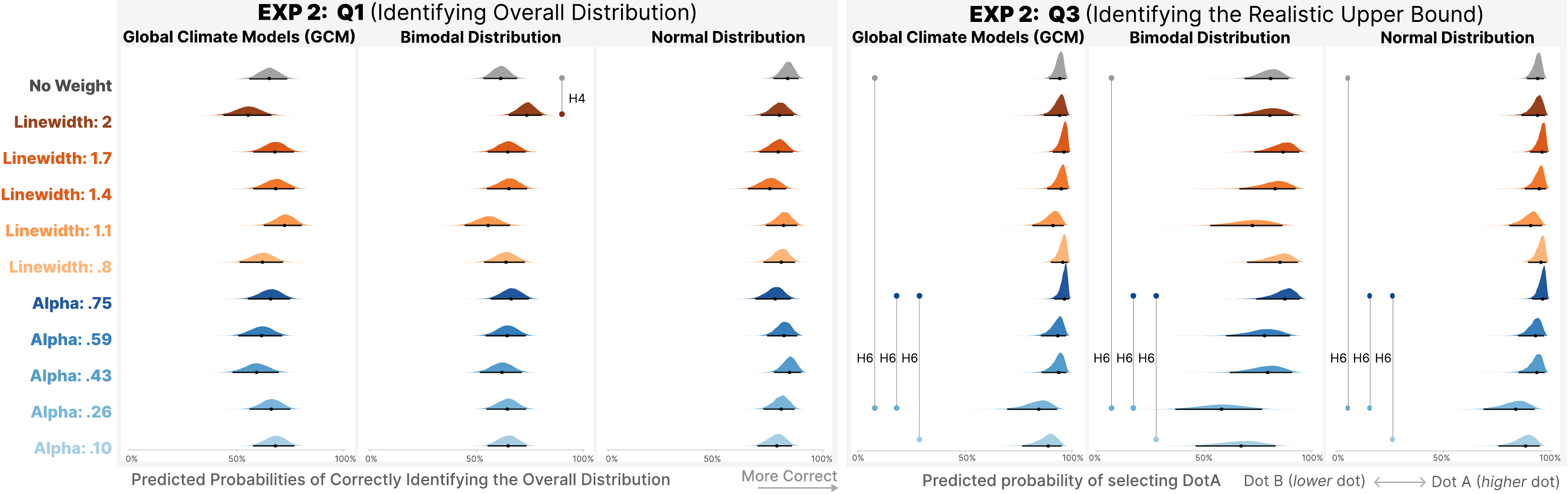}
\caption{
Posterior predicted probability densities for Experiment 2 Q1 (left) and Q3 (right). Each row shows a weighting condition and each column a distribution. Hypotheses annotated on the figure represent meaningfully different contrasts (i.e., the posterior contrast 95\% CrI does not include 0).
}
\label{fig:exp2-results}
\vspace{-8px}
\end{figure*}

\subsection{\textbf{H4}: Weighted forecasts vs. No weighting}

\textbf{H4: At least one weighting condition will produce greater accuracy in identifying the underlying distribution than no weighting.} 
For the Q1 task, one weighting condition showed an improvement. In the Bimodal condition, the strongest linewidth manipulation (\Linewidthtwo) produced a higher predicted probability of selecting the correct response than the no-weighting baseline (\autoref{fig:exp2-results} left panel H4, $\Delta=0.13$, 95\% CrI [0.03, 0.22]). No other alpha or linewidth conditions produced credible improvements relative to the baseline in Q1. For Q2, there was no evidence that weighting improved performance relative to the no-weighting condition. Visual weighting through \Alpha and \Linewidth did not alter readers' ability to perceive the overall distribution.

Overall, these results provide \textbf{no consistent proof for H4}. Weighting did not consistently impact participants' ability to identify the underlying data distribution. It is worth noting that weighting in our study correlates with the statistical distribution of the forecast, meaning that such dual encoding did not impact distribution perception. However, we argue that the weighting's lack of effect on forecast distribution perception may be useful in visualization practice. Visual weighting that does not interfere with distributional perception can be used instead to encode other relevant forecast attributes into MFV.

\subsection{\textbf{H5}: Moderate weighting advantage}
\textbf{H5: Accuracy will be highest at moderate vs. minimal or extreme weighting levels.} 
Overall, the results did not show a consistent moderate weighting advantage. In Q1, the only credible difference emerged in the Bimodal condition, but the difference did not support the hypothesis. Instead, the strongest linewidth weighting (\Linewidthtwo) produced higher accuracy than a moderate linewidth weighting (\Linewidthoneone; $\Delta = 0.18$, 95\% CrI [0.06, 0.30]), and than the non-weighted baseline. In Q2, one credible difference appeared for the Normal condition, where intermediate alpha weighting (\Alphafivenine) produced higher accuracy than the moderate alpha weighting (\Alphafourthree,  $\Delta = 0.09$, 95\% CrI [0.01, 0.19]). No other contrasts excluded zero. These results \textbf{do not support H5}. This pattern suggests that accuracy did not reliably follow our hypothesized inverted-U relationship across weighting strength.

\subsection{\textbf{H6}: Extent effect}
\textbf{H6: When extrema forecasts are shown with faintest lines or thinnest linewidth, participants will tend to choose lower upper‑limit responses.} 
Consistent with this prediction, the strongest evidence emerged for the faintest alpha conditions (\autoref{fig:exp2-results} left panel, H6). Across all three distributions, \Alphatwosix produced lower predicted probabilities of selecting DotA than the non-weighted baseline (e.g., Bimodal: $\Delta = -0.23$, 95\% CrI [-0.43, -0.04]). The \Alphaone condition showed effects in the same direction, but differences were smaller and less certain relative to the baseline (e.g., Bimodal: $\Delta = -0.14$, 95\% CrI [-0.34, 0.03]). Both low-alpha conditions also tended to produce lower probabilities of selecting DotA than the highest alpha condition, \Alphasevenfive, with the clearest differences observed for \Alphatwosix. In contrast, evidence for linewidth manipulations was weaker and less consistent. Contrasts between linewidth levels and the non-weighted baseline were small and uncertain, with credible intervals including zero.

Overall, these results provide \textbf{limited support for H6}, suggesting that dual encoding distributional properties impacts extent perception. Deemphasizing extrema forecasts with low alpha, especially \Alphatwosix, reduced participants' tendency to select the higher-valued upper-limit response, consistent with an extent-based heuristic. However, this pattern was not consistently observed for linewidth manipulations.

\section{Discussion}
\label{sec:discussion}

Our results suggest that MFV may better reflect the underlying forecast distribution than summary-based visualizations. In Experiment 1, we found that participants were more accurate at identifying the underlying forecast distribution when viewing MFV than when viewing CI-based visualizations. In Experiment 2, we provide evidence that modulating forecast rendering (with alpha) can influence extent heuristics with limited influence on the overall perception of forecast distribution.

\subsection{Discussion of Findings}

\noindent \textbf{Distribution Perception.}~~
Our findings comparing the visual perception of MFV and CI-based visualizations (Experiment 1) complement and extend prior works on uncertainty visualization~\cite{sarma2023evaluating, sarma2025more, tak2014perception}. Prior work has shown that different encoding of distributions can affect readers' task performance and introduce trade-offs between bias and precision~\cite{sarma2023evaluating}, and that summary-based visualizations can suppress information about the frequency and arrangement of individual forecasts~\cite{sarma2025more}. Our work provided further evidence of those findings in the context of climate forecast line charts. Additionally, complementing prior work on uncertainty visualization for line charts~\cite{tak2014perception}, we found that CI-based visualizations can encourage the normality perception in non-normal forecast sets. Extending these findings, we show that MFV can better inform the underlying data shape across four simulated distribution types than equivalent CI-based visualizations, with the strongest evidence observed under the full-forecast condition (i.e., MFV22). Because real-world forecasts, such as our test case of climate forecasts, are not always well characterized by normal distributions, MFV may help readers reduce the likelihood of mischaracterizing distributions.

Motivated by a practical consideration in which showing a full forecast set may cause visual clutter, we explored whether and how, through two weighting schemes (linewidth and alpha), a sampled MFV can preserve distributional information relative to the full MFV. While we found no consistent evidence of a desirable impact of weighing, we also found no evidence that it notably hinders the perception of distributions. Such properties outline opportunities in designing distribution-preserving encodings that serve other visual communication goals, such as visualizing models that are not epistemically equivalent~\cite{knutti2017climate} through attributes such as precision, reliability, or representativeness.

\smallskip
\noindent \textbf{Extent Heuristic.}~~Consistent with prior work~\cite{padilla2026examininga}, participants appeared to use the furthest visible extent of a visual mark as a heuristic for plausible upper outcomes, a pattern that has been previously observed for MFV, and that is related to within-the-bar bias documented for standard confidence intervals~\cite{belia2005researchers}. A similar pattern emerged in the faintest alpha conditions of Experiment 2. When the most extreme trajectories were shown with very low opacity, participants were more likely to select inner upper-bound responses, indicating that they treated these de-emphasized forecasts as less plausible. This effect was not driven by participants being unable to see the lines. Only 2\% of participants in Experiment 2 reported that they could not see the faintest two lines, whereas 24\% of participants selected the lower response option. In applied settings such as climate forecasting, this suggests that low-alpha encodings may help keep outlying or lower-priority forecasts visible without encouraging viewers to treat them as equally likely.

\subsection{Limitations and Future Work}

There are several limitations in our study design. First, we intentionally simplified our studies' sampling and weighting schemes. We selected representative forecasts using a deterministic percentile-based procedure and derived weights from a kernel-density estimate at the forecast horizon. While these choices allowed us to isolate perceptual effects, they do not capture all approaches used in forecasting. The sampling method may also imply certain bias that impacts how participants perceive the visualizations. Second, we simulated the Normal, Bimodal, and Uniform distributions based on the GCMs. While we might have maintained GCM's overall shape in the simulated cases, we did not fully reproduce its noise structure. GCM might also contain implicit biases that affect readers' perceptions. Third, the design of the response options can be further improved. For instance, we designed the response option for Q1 to have a similar shape to MFV, which might lead participants to use format-matching and advantage the MFV condition.

Additionally, there are several factors that might impact how our work generalizes. First, although climate forecasting provides a realistic test bed, future work is required to generalize our findings to other contexts. While we accounted for participants' attitudes towards climate change in our analyses, there may also be other individual-difference factors we did not account for. For example, the perceptual effects of weighting may differ across domains, forecast trajectories, and readers' distinct visual abilities and expectations. Furthermore, the weighting was based on forecast consensus and therefore inherently dependent on the forecast distribution. Future work is needed to test whether these effects generalize when weights encode independent forecast attributes, such as precision, reliability, or representativeness. Second, our tasks focused on distribution perception and upper-limit judgment. Future work should examine how weighted MFV affects downstream decisions, trust, perceived credibility, and willingness to act on forecasts. Last, our participants were limited to the lay U.S. population and may not generalize to other countries or trained experts.

\section{Conclusion}
\label{sec:conclusion}

In this work, we show that MFV can convey distributional information more accurately than CIs. We reinforce MFV's central affordance that they preserve perceptual information (e.g., shape, spread) of multiple forecasts better than summary-based alternatives. At the same time, we show that MFV can be potentially extended to communicate model-level differences through visual weighting without compromising perception of broader forecast distribution. We argue that this is an especially important attribute for modern forecasting practice, which increasingly requires not only presenting multiple forecasts but also conveying meaningful differences among them. In summary, these findings position MFV not just as an alternative to confidence intervals but as a flexible foundation for richer, more faithful forecast communication adaptable to many real-world modeling workflows.

\section*{Supplemental Materials}
\label{sec:supplemental_materials}

All supplemental materials are publicly available on OSF at  \href{https://osf.io/fx67m/}{\texttt{osf\discretionary{}{.}{.}io\discretionary{/}{}{/}fx67m}}, released under a \href{https://creativecommons.org/licenses/by/4.0/}{\ccby{} CC BY 4.0 license}. In particular, they include (1) the interview script used for the need finding study, (2) preregistration of both Experiments, (3) the data/script for generating the stimuli and the stimulus/response options used in the study, and (4) the data and script for the Bayesian analysis.

\acknowledgments{%
This work was supported in part by NSF Grant \#2403094, NIH Grant \#1R01AI188576-01, and a Northeastern University Tier-1 Research Seed Grant. We want to thank all participants for taking part in our study. We also extend our thanks to the anonymous reviewers for their constructive feedback.
}

\balance

\bibliographystyle{abbrv-doi-hyperref}
\bibliography{bib/additions.bib, bib/export.bib}

\newpage
\appendix %
\crefalias{section}{appendix} %

\section{Parameters for Generating Data Distributions}
\label{appendix:data-dist-parameters}

We generated simulated distributions (Uniform, Normal, Bimodal) based on the GCM data at four anchor years $a \in \{2026, 2050, 2075, 2099\}$ (forecast values at anchor year $a$ are denoted $g(a)$). The 2026 anchor used the original GCM values (i.e., $g(2026)$). For the rest anchor years (i.e., $a' \in \{ 2050, 2075, 2099\}$), we generated values from the corresponding GCM summary statistics (e.g., with operators \(\min\), \(\max\), \(\operatorname{median}\), and \(\operatorname{sd}\)).

\begin{table}[h]
  \centering
  \caption{Parameters used to generate synthetic data distributions.}
  \label{tab:synthetic-distribution-parameters}
  \small
  \begin{tabular}{@{}llccc@{}}
    \hline
    \textbf{Distribution} & \textbf{Scenario} & \textbf{Center} & \textbf{Spread (SD)} & \textbf{Shift} \\
    \hline
    Uniform & All & \multicolumn{2}{c}{\([\min(g(a')), \max(g(a'))]\)} & -- \\
    Normal & All & \(\operatorname{median}(g(a'))\) & \(\operatorname{sd}(g(a'))\) & -- \\
    \multirow[c]{3}{*}{Bimodal} & SSP1-2.6 & \(\operatorname{median}(g(a'))\) & \(\sigma(a'): 1.6 \rightarrow 0.9\) & \(s(a'): 0 \rightarrow 1.8\) \\
    & SSP2-4.5 & \(\operatorname{median}(g(a'))\) & \(\sigma(a'): 1.6 \rightarrow 0.9\) & \(s(a'): 0 \rightarrow 1.8\) \\
    & SSP5-8.5 & \(\operatorname{median}(g(a'))\) & \(\sigma(a'): 1.4 \rightarrow 1.5\) & \(s(a'): 0 \rightarrow 2.5\) \\
    \hline
  \end{tabular}
\end{table}

We document the parameters used to generate data for each distribution-scenario pair in \autoref{tab:synthetic-distribution-parameters}. $\sigma$ is a linear function that defines the standard deviation for each year, and $s$ is an additive shift function. Both $\sigma$ and $s$ were defined over the full anchor-year sequence $a$, but values were only evaluated at \(a'\). Specifically for the simulated Bimodal distribution, the shift parameter \(s(a')\) is the one-sided distance from the GCM median to each component's mean, i.e., the two component centers are \(\operatorname{median}(g(a')) \pm s(a')\), with spread $\sigma(a')$.

\begin{table*}[t]
\centering
\caption{
Summary of hypothesis outcomes across Experiments~1 and~2.
For each hypothesis, we report the questions, distributions, and pairwise contrasts that provided evidence. We also report the total contrasts, interpretation, and the count/proportions of three categories: credible positive (95\% CrI excludes 0 in the predicted direction), directionally positive but uncertain ($\Delta > 0$), or not supported.
}
\label{tab:hypotheses-summary}
\scriptsize
\renewcommand{\arraystretch}{1.18}

\newcolumntype{Y}{>{\raggedright\arraybackslash}X}
\newcolumntype{C}{>{\centering\arraybackslash}p{0.03\textwidth}}
\newcolumntype{H}{>{\centering\arraybackslash}p{0.02\textwidth}}
\newcolumntype{Q}{>{\centering\arraybackslash}p{0.03\textwidth}}
\newcolumntype{P}{>{\arraybackslash}p{0.19\textwidth}}
\newcolumntype{D}{>{\raggedright\arraybackslash}p{0.12\textwidth}}

\newcommand{\VsRef}[2]{(#1) $-$\par (#2)}

\begin{tabularx}{\textwidth}{@{}
  H
  Y
  Q
  D
  P
  C
  C
  C
  C
  p{0.05\textwidth}
@{}}
\toprule
\textbf{ID} &
\textbf{Hypothesis} &
\textbf{Task(s)} &
\textbf{Dist.(s)} &
\textbf{Pairs} &
\textbf{Total Comp(s).} &
\textbf{Cred. pos.} &
\textbf{Dir. pos.} &
\textbf{Not supp.} &
\textbf{Interp.} \\
\midrule

\multicolumn{10}{@{}l}{\textbf{Experiment 1}} \\
\midrule

H1A
& \MFVfull improves distribution-identification accuracy relative to \CI-based visualizations.
& Q1, Q2
& Bimodal, GCM, Normal, Uniform
& \VsRef{\MFVfull}{\CIfull, \CIninefive}
& 16 & 11 (.69) & 5 (.31) & 0 (.00)
& Strong Support \\

H1B
& \MFVfull improves accuracy relative to \MFVnine.
& Q1, Q2
& Bimodal, GCM, Normal, Uniform
& \VsRef{\MFVfull}{\MFVnine}
& 8 & 4 (.50) & 2 (.25) & 2 (.25)
& Limited support \\

H1C
& \MFVnine improves accuracy relative to \CI-based visualizations.
& Q1, Q2
& Bimodal, GCM, Normal, Uniform
& \VsRef{\MFVnine}{\CIfull, \CIninefive}
& 16 & 9 (.56) & 5 (.31) & 2 (.13)
& Moderate support \\

H2A
& For non-Normal distributions, \CIfull elicits more Normal responses than \MFV.
& Q1, Q2
& Bimodal, GCM, Uniform
& \VsRef{\CIfull}{\MFVfull, \MFVnine}
& 12 & 10 (.83) & 2 (.17) & 0 (.00)
& Strong Support \\

H2B
& For non-Normal distributions, \CIninefive elicits more Normal responses than \MFV.
& Q1, Q2
& Bimodal, GCM, Uniform
& \VsRef{\CIninefive}{\MFVfull, \MFVnine}
& 12 & 6 (.50) & 5 (.42) & 1 (.08)
& Moderate support \\

H2C
& For Normal distributions, \CIfull elicits similar Normal-response rates as \MFV.
& Q1, Q2
& Normal
& \VsRef{\CIfull}{\MFVfull, \MFVnine}
& 4 & 0 (.00) & 0 (.00) & 4 (1.00)
& Not supported$^{\ast}$ \\

H2D
& For Normal distributions, \CIninefive elicits similar Normal-response rates as \MFV.
& Q1, Q2
& Normal
& \VsRef{\CIninefive}{\MFVfull, \MFVnine}
& 4 & 0 (.00) & 0 (.00) & 4 (1.00)
& Not supported$^{\ast}$ \\

H3
& \CI-based visualizations elicit lower-valued upper-limit responses than \MFV.
& Q3
& Bimodal, GCM, Normal, Uniform
& \VsRef{\MFVfull, \MFVnine}{\CIfull, \CIninefive}
& 16 & 16 (1.00) & 0 (.00) & 0 (.00)
& Strong Support \\

\midrule

\multicolumn{10}{@{}l}{\textbf{Experiment 2}} \\
\midrule

H4
& At least one weighting condition improves distribution-identification accuracy relative to no weighting.
& Q1, Q2
& Bimodal, GCM, Normal
& \VsRef{\Alphaone, \Alphatwosix, \Alphafourthree, \Alphafivenine, \Alphasevenfive;
    \Linewidtheight, \Linewidthoneone, \Linewidthonefour, \Linewidthoneseven, \Linewidthtwo
  }{no weighting}
& 60 & 1 (.02) & 20 (.33) & 39 (.65)
& Not supported \\

H5
& Accuracy is highest at moderate weighting levels compared with minimal or extreme weighting levels.
& Q1, Q2
& Bimodal, GCM, Normal
& \VsRef{\Alphafourthree
  }{\Alphaone, \Alphatwosix, \Alphafivenine, \Alphasevenfive
  }\par
  \VsRef{\Linewidthoneone}{\Linewidtheight, \Linewidthonefour, \Linewidthoneseven, \Linewidthtwo}
& 48 & 2 (.04) & 25 (.52) & 21 (.44)
& Not supported \\

H6
& Deemphasizing extrema forecasts with faintest alpha or thinnest linewidth reduces higher upper-limit responses.
& Q3
& Bimodal, GCM, Normal
& \VsRef{no weighting, \Alphasevenfive}{\Alphaone, \Alphatwosix} \par
\VsRef{no weighting, \Linewidthtwo}{\Linewidtheight, \Linewidthoneone}
& 24 & 9 (.38) & 9 (.38) & 6 (.25)
& Limited support \\

\bottomrule

\addlinespace[2pt]
\multicolumn{10}{@{}p{\textwidth}@{}}{
\scriptsize \textit{$^{\ast}$H2C-H2D predicted similarity rather than a directional difference. Thus, ``not supported'' in those hypotheses indicates that the contrasts did not provide evidence for similar Normal-response rates; instead, observed contrasts generally favored \MFV over the equivalent \CI display.}
} \\
\addlinespace[2pt]

\end{tabularx}
\end{table*}

We generated values between the anchor years using linear interpolation with Gaussian noise. The standard deviation of the added Gaussian noise increased linearly from 0.02 to 0.08 across 2026 to 2099. Noise was set to 0 at the anchor years.

\section{Rationale for Alpha and Linewidth Levels}
\label{appendix:alpha-linewidth-levels}
Assuming a 24-inch viewing distance and a stimulus size of 2048 $\times$ 1463 px at 300 dpi, our stimuli would be approximately 173.48 mm wide, corresponding to a horizontal visual angle of approximately $16.3^\circ$. Under this geometry, our thinnest stroke width (0.5 mm) subtends approximately $0.047^\circ$ of visual angle, and our smallest adjacent stroke width step (0.5 mm to 0.8 mm) differs by approximately $0.028^\circ$. We selected this minimum step to avoid very small differences while preserving a realistic range of visual weighting levels. Although prior work does not directly evaluate alpha transparency, luminance is likely the closest analogue in our setting. Sterzik et al.~\cite{sterzik2024perception} suggest that luminance differences of 10\% and 20\% can fall below the discrimination threshold in some regions of the scale, whereas larger differences are more clearly perceptible. We therefore selected a minimum alpha step of 0.25. Although the luminance guideline can only approximate alpha in our context, it provides a conservative justification for designing our smallest transparency manipulation so that it is perceptually discriminable. After selecting both channels' upper/lower bounds and minimum distance, we used equal intervals to determine the remaining linewidth and alpha step differences.

\section{Summary of Hypothesis and Outcome}
\label{appendix:hypo-outcome}

We report our hypothesis, comparison structure, evidence, and judgments of our analysis in \autoref{tab:hypotheses-summary}.

We interpreted the evidence for each hypothesis based on the pattern of pairwise contrasts. Specifically, we classified each contrast as \textit{credible positive} when the 95\% credible interval excluded zero in the predicted direction, \textit{directionally supportive but uncertain} when the estimated contrast was in the predicted direction, but the 95\% credible interval included zero, and \textit{not supportive} otherwise. We then summarized each hypothesis as having \textit{strong support}, \textit{moderate support}, \textit{limited support}, or being \textit{not supported}, based on the consistency and strength of evidence across its tested contrasts.

\end{document}